# The Low-Radioactivity Underground Argon Workshop
A workshop synopsis

Thomas Alexander[1], Henning O. Back[1,*], Walter Bonivento[2], Mark Boulay[3], Philippe Collon[4], Zhongyi Feng[5], Michael Foxe[1], Pablo García Abia[6], Pietro Giampa[7], Christopher Jackson[1], Christine Johnson[1], Emily Mace[1], Peter Mueller[8], László Palcsu[9], Walter Pettus[10], Roland Purtschert[11], Andrew Renshaw[12], Richard Saldanha[1], Kate Scholberg[13], Marino Simeone[14], Ondřej Šrámek[15], Rex Tayloe[16], Ward TeGrotenhuis[1], Signe White[1], Richard Williams[1]

1. Pacific Northwest National Laboratory, Richland, Washington 99352, USA
2. INFN Cagliari, Cagliari 09042, Italy
3. Carleton University, Ottawa, Ontario K1S 5B6, Canada
4. University of Notre Dame, Notre Dame, Indiana 46556, USA
5. Kirchhoff-Institut für Physik, Universität Heidelberg, 69120 Heidelberg, Germany
6. CIEMAT, Centro de Investigaciones Energéticas, Medioambientales y Tecnológicas, Madrid 28040, Spain
7. TRIUMF, Vancouver, British Columbia V6T 2A3, Canada
8. Argonne National Laboratory, Lemont, Illinois 60439, USA
9. Institute for Nuclear Research, Hungarian Academy of Sciences, H-4001 Debrecen, Hungary
10. University of Washington, Seattle, Washington 98195, USA
11. University of Bern, 3012 Bern, Switzerland
12. University of Houston, Houston, Texas 77204, USA
13. Duke University, Durham, North Carolina 27708, USA
14. Università degli Studi di Napoli "Federico II", Napoli 80125, Italy
15. Department of Geophysics, Faculty of Mathematics and Physics, Charles University, 180000 Prague, Czech Republic
16. Indiana University, Bloomington, Indiana 47405, USA

## Abstract

In response to the growing need for low-radioactivity argon, community experts and interested parties came together for a 2-day workshop to discuss the worldwide low-radioactivity argon needs and the challenges associated with its production and characterization. Several topics were covered: experimental needs and requirements for low-radioactivity argon, the sources of low-radioactivity argon and its production, how long-lived argon radionuclides are created in nature, measuring argon radionuclides, and other applicable topics. The Low-Radioactivity Underground Argon (LRUA) workshop took place on March 19-20, 2018 at Pacific Northwest National Laboratory in Richland Washington, USA. This paper is a synopsis of the workshop with the associated abstracts from the talks.

---

*Corresponding author. Present address: Pacific Northwest National Laboratory, 902 Battelle Boulevard, P.O. Box 999, MSIN J4-65, Richland, WA 99352 USA*
*E-mail address: henning.back@pnnl.gov (H. O. Back)*

# Table of Contents

# 1 Introduction

Argon derived from the atmosphere, which is predominately stable $^{40}$Ar, contains cosmogenically produced long-lived radioactive isotopes of argon— $^{42}$Ar, $^{39}$Ar, and $^{37}$Ar. These isotopes are at high enough concentrations in the atmosphere to be a significant background for low-background argon-based radiation detectors, and because all commercial argon is produced from air, these argon radionuclides can represent irreducible backgrounds. However, after a 5-year campaign of extracting and purifying argon from deep $CO_2$ wells in Southwestern Colorado in the United States, the DarkSide-50 dark matter search experiment demonstrated that this unique argon contained an $^{39}$Ar concentration 0.073% of that in the atmosphere, or approximately 0.73 mBq/kg$_{Ar}$ [1]. Cosmogenically produced $^{37}$Ar was also detected in the early running of the DarkSide-50 detector [1], but $^{42}$Ar was not observed. This demonstrated that large-mass quantities of low-radioactivity underground argon can be obtained and has sparked a global interest in a sustained supply of low-radioactivity underground argon to meet the needs of a broad range of disciplines from nuclear and particle physics to environmental studies and national security.

The broader availability of low-radioactivity argon and the challenges associated with finding new sources, with current and future production of underground argon, and with the characterization of the argon itself have become important topics associated with the growing need. As an example, measuring the concentration of $^{39}$Ar in underground argon is critical to both searching for new sources and as quality assurance and quality control during production. Additionally, the determination of argon radionuclide production underground and in the atmosphere is relevant to both new underground argon source characterization and in understanding cosmogenic activation when the underground argon resides close to or on the surface of the Earth. Both issues are also relevant in the context of groundwater dating using $^{39}$Ar. Another consideration is whether isotope separation is required to achieve the level of argon-radioisotope depletion needed for some applications.

In response to this growing need for low-radioactivity argon and the open questions about new sources, production, and characterization, the 2-day Low-Radioactivity Underground Argon (LRUA) workshop was held at Pacific Northwest National Laboratory in Richland, WA, on March 19-20, 2018. The LRUA workshop brought together a diverse interdisciplinary group of scientists and engineers, community experts, and interested parties to discuss the worldwide low-radioactivity argon needs and the challenges associated with its production and characterization.

This paper details the needs for low-radioactivity underground argon that were presented at the workshop, and provides an introduction to the other workshop topics and the full set of talk abstracts. Links to the full presentations are provided in the references with associated DOI numbers.

# 2 The global needs for low-radioactivity underground argon

The largest needs for low-radioactivity underground argon are in the fundamental nuclear and particle physics fields. The DarkSide experiments have been driving the demand and production for low-radioactivity underground argon [1], but with that success the demands have risen. Beyond WIMP dark matter detection, the physics that is more easily reached by the availability of low-radioactivity underground argon includes: neutrinoless double-beta decay by eliminating $^{42}$Ar and $^{39}$Ar in the argon that surrounds the germanium crystals of the LEGEND experiment [2], measuring low-energy neutrinos in the DUNE detector by reducing the $^{39}$Ar beta rate and also the higher-energy beta from $^{42}$K (the daughter of $^{42}$Ar) [3], and coherent elastic neutrino-nucleus scattering within the series of COHERENT experiments through increasing live time by reducing $^{39}$Ar decays [4].

The potential needs for low-radioactivity underground argon span from tens of kilograms for the COHERENT experiment to tens of kilotonnes for the DUNE modules. Likewise, the requirements for $^{39}$Ar reduction from the atmospheric concentration span several orders of magnitude from a factor of 10 reduction for COHERENT to more than a factor of 1000 for future argon-based dark matter searches. The specific needs for low-radioactivity underground argon are reviewed for five distinct use cases in the remainder of this section.

## A Unified Program of Argon Dark Matter Searches: DarkSide-50, DarkSide-20k and The Global Argon Dark Matter Collaboration [5]
*Walter Bonivento – INFN Cagliari (for the DarkSide collaboration)*

In 2018 The DarkSide-50 experiment at LNGS produced two world leading results in the search for WIMP dark matter with spin-independent interactions, i.e. the best world limit between 1.5 and 6 GeV/c$^2$ of mass with the ionization-only signal analysis [6] and the best world limit with argon detectors for high mass above 30 GeV/c$^2$ [7]. Researchers from the ArDM experiment at LSC; the DarkSide-50 experiment at LNGS; the DEAP-3600 experiment at SNOLAB; and the MiniCLEAN experiment at SNOLAB will jointly carry out, as the single next step for the large mass WIMP searches with argon at the scale of a few tens of tonnes, the DarkSide-20k experiment at Laboratori Nazionali del Gran Sasso.

DarkSide-20k is a 20-tonne fiducial volume dual-phase TPC to be operated at LNGS with 37 tonnes of underground argon (UAr), designed to collect an exposure of 100 tonne×years, completely free of neutron-induced nuclear recoil backgrounds and all electron recoil backgrounds. DarkSide-20k is set to start operating by 2023 and will have sensitivity to WIMP-nucleon spin-independent cross sections of $1.2 \times 10^{-47}$ cm$^2$ for WIMPs of 1 TeV/c$^2$ mass, to be achieved during a 5-year run. An extended 10-year run could produce an exposure of 200 tonne×years, with sensitivity for the cross section of $7.4 \times 10^{-48}$ cm$^2$, for the same WIMP mass. DarkSide-20k will explore the WIMP-nucleon cross-section down to the edge of the "neutrino floor", where coherent neutrino-nucleus scattering from environmental neutrinos induce nuclear recoils in the detector.

At the same time, the DarkSide collaboration will pursue the search for low-mass WIMPs with a 1-tonne detector, DarkSide-LowMass, aiming also at reaching the "neutrino floor" for masses below 8 GeV/c$^2$. The level of $^{39}$Ar depletion impacts directly the sensitivity of DarkSide-LowMass. We therefore plan to operate DarkSide-20k with UAr processed for further reduction of $^{39}$Ar with the Aria plant in Sardinia, the tallest cryogenic distillation plant ever conceived [8].

The combination of DarkSide-20k and DarkSide-LowMass will require the extraction and purification of an overall amount of 50 tonnes of underground argon.

## DEAP-3600 and discussion of a multi-hundred tonne argon detector for dark matter [9]
*Mark Boulay – Carleton University (for the DEAP and Global Argon Dark Matter Colaborations)*

DEAP-3600 is a novel experiment searching for dark matter particle interactions on several tonnes of liquid argon at SNOLAB. The argon is contained in a large ultralow-background acrylic vessel viewed by 255 8-inch photomultiplier tubes. The detector has been designed to allow control of radioactive backgrounds for an ultimate sensitivity to spin-independent scattering of $10^{-46}$ cm$^2$ per nucleon at 100 GeV mass, and has been collecting data with atmospheric argon since late 2016. Although DEAP-3600 does not require low-radioactivity argon to reach its target sensitivity, target masses beyond the DEAP-3600 mass necessitate the use of underground argon. Underground argon in DEAP-3600 would allow a lower analysis threshold, increasing the experimental sensitivity.

Plans are being developed within the Global Argon Dark Matter Collaboration for an experiment beyond DarkSide-20k. This will be a multi-hundred tonne detector that will approach the neutrino floor to produce a 1,000 tonne-year exposure. The multi-hundred tonne detector will require upwards of 500 tonnes of low-radioactivity underground argon. The level of $^{39}$Ar will depend on the technology chosen for this detector, however, a single-phase detector like DEAP-3600 can likely run with the $^{39}$Ar concentration found in the DarkSide-50 underground argon. The multi-hundred tonne detector is projected to begin its dark matter search in 2027, which is when the depleted underground argon target would be needed.

## Low-Energy Neutrinos in DUNE [3]
*Kate Scholberg – Duke University (for the DUNE Collaboration)*

A core-collapse supernova in the Milky Way will create an enormous burst of neutrinos of all flavors with a few tens of MeV, within about ten seconds, observable in detectors around the world [10]. The 70-kton (40-kton fiducial) Deep Underground Neutrino Experiment (DUNE) liquid argon time projection chamber detectors will have unique sensitivity to the $\nu_e$ component of this burst, via the dominant charged-current (CC) interaction $\nu_e + {}^{40}\text{Ar} \rightarrow e^- + {}^{40}\text{K}^*$ channel [11, 12]. In principle, presence or absence of characteristic final-state gamma deexcitation cascades can help to tag low-energy-neutrino interaction channels in DUNE, in order to disentangle $\nu_e$ CC from subdominant neutral-current deexcitations and elastic scattering on electrons. In addition, steady-state signals from solar neutrinos and diffuse supernova background neutrinos are in principle observable, but have much more challenging background and triggering issues.

Although a deep underground location is highly beneficial for low-energy neutrino detection in DUNE, long-lived radiologicals in argon can still create problems. $^{39}$Ar is the dominant radiological; although single decays are not likely to create more than single-wire blips, the accumulation of these in the ~m$^3$ volume occupied by a supernova event can potentially degrade the reconstruction by faking signals from Compton scatters of nuclear deexcitation or bremsstrahlung γ's. More seriously, the radiological signal creates a large data rate, and a challenge for data acquisition/trigger design. The DAQ must maintain good supernova burst and individual-event reconstruction efficiency, while satisfying a requirement of radiological-dominated data volume of less than 1 PB per year per 10-kton detector. Light from $^{39}$Ar decays also affects optical photon detection. For these reasons, low-radioactivity argon would be desirable for DUNE (although of course very large quantities would be needed!)

On the positive side, it is conceivable that $^{39}$Ar decays, thanks to a well-known decay spectrum and uniform distribution within the detector, could be used as a calibration source by providing monitors of electron lifetime and spatial/temporal variations [13].

## CEvNS with a liquid argon scintillation detector [4]
Rex Tayloe – Indiana University (for the COHERENT collaboration)

The COHERENT collaboration is deploying a suite of low-energy detectors in a low-background corridor of the ORNL Spallation Neutron Source (SNS) to measure coherent elastic neutrino nucleus scattering (CEvNS) on an array of nuclear targets employing different technologies [14]. A measurement of CEvNS on different nuclei will test the $N^2$-dependence of the CEvNS cross section and further the physics reach of the COHERENT effort. The first step of this program has been realized with the observation of CEvNS in a 14.6 kg CsI detector. A 22 kg, single-phase, LAr detector (CENNS-10) started data-taking in Dec. 2016 and will provide results on CEvNS from a much lighter nucleus. One year of running with this detector will provide adequate events for verification of the $N^2$-dependence at the standard-model CEvNS cross section.

The CEvNS events of interest in argon have energy in range 20-200 keV. The liquid argon employed in the CENNS-10 detector is high-purity for $N_2$, $O_2$, and $H_2O$ but with no specification on the $^{39}$Ar content. The rate of β-decay (with endpoint of 565 keV) has been measured, in-situ, to be consistent with the canonical 1 Bq/kg. The neutrino source is pulsed with an effective duty factor of about $10^{-5}$, so an SNS-year of neutrino data may be collected in 20 minutes of detector livetime. This low duty-factor, combined with pulse-shape discrimination, allows for a manageable rate of $^{39}$Ar in the CEvNS region of interest and a background subtraction with expected ≈10% statistical precision. However, a reduced rate of $^{39}$Ar will reduce statistical errors that arise in the subtraction of steady state backgrounds, especially in large detectors desiring higher precision.

A ton-scale CEvNS detector is currently being proposed and could be realized within the next 2–3 years. This detector will provide several thousand CEvNS events per year and would benefit from an argon target with a reduced level of $^{39}$Ar. A reduction in $^{39}$Ar by a factor of 10-100×, as has been observed in some underground argon sources, would improve this liquid argon CEvNS program significantly.

## The LAr veto for the Large Enriched Germanium Experiment for Neutrinoless Double Beta Decay (LEGEND) [2]
Walter Petus – University of Washington (for the LEGEND collaboration)

The LEGEND collaboration is pursuing a tonne-scale $^{76}$Ge neutrinoless double-beta decay (0νββ) experiment with discovery potential at a half-life significantly longer than $10^{27}$ years [15]. Observation of 0νββ would violate lepton number conservation, establish the Majorana nature of neutrinos, and serve as an indirect probe of the absolute neutrino mass scale. LEGEND will build on the successes of the current generation of $^{76}$Ge 0νββ experiments, GERDA [16] and the MAJORANA DEMONSTRATOR [17], to limit the background to the level of <0.1 counts/FWHM/t/yr in the region of interest ($Q_{ββ}$ = 2039 keV).

Central to background suppression is an active liquid argon (LAr) veto, which relies on collection of the 125 nm argon scintillation photons. This tags a complementary population of Compton scatter gammas to that identified by the Ge-detector pulse-shape analysis (PSA). Background from $^{42}$Ar-chain decays originating within the LAr veto is intensified as the $^{42}$K daughter ions drift in the field of the Ge detectors towards the array; the subsequent beta decay ($Q_β$ = 3.5 MeV) may penetrate the ~1 mm detector dead layer. This background is projected to contribute near $Q_{ββ}$ up to the full background budget of the tonne-scale experiment without mitigation. The background from $^{39}$Ar also impacts the low-energy physics reach of the experiment by dominating the spectrum below 500 keV.

Both argon-related backgrounds can be functionally eliminated by using underground argon. A segregated argon veto is proposed where the Ge detectors are split into modules (*e.g.,* four modules with 250 kg of Ge detectors in each), which are independently deployed into low radioactivity underground argon volumes. The total required low radioactivity underground argon is ~21 tons (15 m$^3$). The timeline for LEGEND-1000 is subject to project funding, but it would not start before the predecessor LEGEND-200 completes its multi-year data campaign, which begins in 2021.

## Low-Radioactivity Underground Argon and environmental measurements
Henning Back – Pacific Northwest National Laboratory

Environmental studies that use the long-lived argon radionuclides, such as measuring low-levels of $^{39}$Ar for age-dating ground water and $^{37}$Ar detection for nuclear explosion monitoring, have a need for an enduring supply of low-radioactivity underground argon. This background-free argon is used as make-up gas in the detectors if the argon sample is not large enough. Sample carry-over is a potential source for measurement error, which could be eliminated if low-radioactivity underground argon were more readily available and could be used to flush detectors and gas handling systems with background-free argon. Currently, underground argon is a preciously scarce resource that is of limited supply. The minor stockpile that exists at PNNL is the only supply outside of the DarkSide-50 detector, and it will soon be exhausted.

It has been estimated that access to 100 kg/year for many years (i.e., decades) would satisfy both current and expanding needs in this area. Although this could be met by stockpiling from the UAr production for fundamental physics experiments, a truly enduring supply would be beneficial as underground storage becomes necessary for the long-term, and institutional knowledge about how the stockpile was obtained may be lost over time. It is not likely that an $^{39}$Ar concentration better than that attained in the DarkSide-50 experiment would be required.

# 3 LRUA Workshop Book of Abstracts

The Low-Radioactivity Underground Argon workshop brought together not only the physicists from the experiments that need low-radioactivity underground argon, but also scientists and engineers from a broad spectrum of fields with other applicable interests in low-radioactivity underground argon. Presentations were given about how underground argon is obtained, how predictions of radioactive argon isotope production are used to search for new sources and characterize current sources of low-radioactivity and to understand the ingrowth of argon radionuclides in the underground argon, how argon radionuclides are used in environmental studies, techniques for measuring low-levels of $^{39}$Ar, and other applicable areas.

What follows is the book of abstracts for the workshop organized by topical area and with a brief description of the applicability of the topics to low-radioactivity underground argon. The topical areas are:

- Low-Radioactivity Underground Argon Sources, Production, and Isotope Separation
- Production of Argon Radioisotopes in the Environment
- Uses of Argon Radioisotopes in the Environment
- Measuring Low-Levels of Argon Radioisotopes
- High-Level $^{39}$Ar Applications

The workshop presentations associated with the abstracts can be found in the ZENODO repository and are referenced for each abstract.

## 3.1 Low-Radioactivity Underground Argon Sources, Production, and Isotope Separation

After a 5-year campaign of extracting argon from underground $CO_2$ wells, the DarkSide-50 experiment showed that the argon from that source was significantly reduced in $^{39}$Ar [1]. This enabled ground breaking results from the DarkSide-50 experiment [18, 6], and has made it possible to build the DarkSide-20k experiment [19]. DarkSide intends to expand their production at the original site to produce their dark matter WIMP target, and potentially for a future 300-tonne scale argon based dark matter WIMP detector [20]. A new source for low-radioactivity underground argon is also being explored in Hungary from wells that are very similar to the DarkSide source [21]. Large-scale active depletion of $^{39}$Ar becomes required if the underground argon is not low enough in $^{39}$Ar, which is being explored with the Aria column in Italy using cryogenic distillation as the isotope separation technology [8]. Modern technology has the potential for drastically reducing the size of these cryogenic distillation columns [22]. As the demand for low-radioactivity underground argon and the required $^{39}$Ar depletion factor grow, new sources will be needed as well as techniques for further reducing the argon radionuclide concentration.

### Developing the First Supply of Underground Argon [23]
Thomas Alexander – Pacific Northwest National Laboratory

The Darkside-50 target was produced as the first application of underground argon. Developing and delivering the target from the ground to the detector took substantial effort. In this talk we will chronicle the efforts taken in the development of the DarkSide-50 target, describing how the source was located, how the gas was purified, and finally how the target was delivered and inserted into the DarkSide-50 detector. The performance and residual $^{39}$Ar content will also be discussed.

### Urania: Extraction of UAr for DarkSide-20k [20]
Andrew Renshaw – University of Houston
(Presented by Henning Back – Pacific Northwest National Laboratory)

The DarkSide-50 (DS-50) two-phase liquid argon (LAr) detector has been searching for weakly interacting massive particle (WIMP) dark matter for the past four years, and during the last three years has been successfully operating the detector with argon that was extracted from underground $CO_2$ wells in Colorado, USA. This source of argon has been long shielded from cosmic rays entering Earth's atmosphere and has been proven to have a lower concentration of the cosmogenically produced isotope of $^{39}$Ar that beta decays with an endpoint energy that causes the beta spectrum to entirely cover the LAr WIMP search region. A 70-day exposure of the underground argon (UAr) inside DS-50 demonstrated that the UAr extracted from Colorado contains a factor greater than 1400 less $^{39}$Ar, compared to argon separated from the air in Earth's atmosphere. This large reduction in $^{39}$Ar opens the door for the construction of much larger two-phase LAr detectors that can be used for the direct detection of WIMP dark matter, as well as other rare-event searches. This talk will focus on the details of a new project called Urania, aiming to extract up to 250 kg/day of UAr from the same source of gas used to extract the UAr for DS-50. This effort will procure 50-tonnes of UAr for use in a 20-tonne fiducial volume detector called DarkSide-20k, which is set to begin operations in 2022.

### A potential low-level argon source: a multistacked $CO_2$ field in Central Europe [21]
László Palcsu - Hungarian Academy of Sciences

In the sedimentary environments of the Pannonian Basin (PB) in central Europe, $CO_2$ filled layers are located in many sites. In addition to that research of $CO_2$ fields has an emphasized importance, such as a natural analogue of the capture and sequestration the industrial $CO_2$, the gas fields contain primordial and ancient trace gases. Analyzing these components we can better understand the processes during the formation of the Earth. Our recent study has shown that the $CO_2$ from the gas fields near Répcelak is originated from the degassing of the upper-mantle and Pliocene intrusions. This $CO_2$ is ascending and being mixed with nitrogen and hydrocarbon gases produced from the sedimentary organic matter, and then it is stacked beneath the impermeable layers. Based on the helium isotope ratios of the gases it has been found that the 25-60% of He is of mantle-derived and the remaining He component is added during the residence time in the reservoir rock and the up-flow of $CO_2$. The amount of radiogenic argon in the gases, as the high $^{40}$Ar/$^{36}$Ar showed, refers to the age of gas fields. The argon is expected to be very old. One goal of our research is to investigate how this argon could be used in promising dark matter research.

### The ARIA project: production of depleted argon for the DarkSide experiment [8]
Marino Simeone – University of Naples

The DarkSide collaboration has been working on the design of detectors operated with argon depleted from $^{39}$Ar. It has been found that the underground argon source from the Cortez $CO_2$ wells located in Colorado is characterized by a specific content of radioactivity from $^{39}$Ar, which is at least 1000 times lower than that of atmospheric argon. However, for detectors specialized for the discovery of low-mass dark matter through the observation of the sole ionization signal, another depletion factor of 10–100 is needed. The ARIA project was developed within the DarkSide collaboration with the purpose of further depleting the underground argon by means of a proper technology at a reasonable cost and in a reasonable time. Isotopic separation by cryogenic distillation was found to be the proper one. Therefore, the ARIA project involves an international team that has been working on the design and installation of a cryogenic distillation plant that allows to perform the isotopic separation of argon. This is a challenging task due to the very tiny difference in the vapour pressure of the species to be separated, which requires a large number of equilibrium stages and, thus, a very tall column for the desired separation to be accomplished. This fact led the Monte Sinni mine, located in Sardinia (Italy), to be identified as the ideal place for the realization of the ARIA project.

The ARIA project also finds application to other fields (e.g., medical and nuclear), where the production of stable isotopes other than those of argon is of paramount importance. Preliminary studies on the production of $^{15}$N and on the production of $^{13}$C using, respectively, nitric oxide and carbon monoxide as operating substances have been carried out by means of classical chemical engineering design methods. The aim of this work is to present the ARIA project and the studies that have been carried out about isotope separation by cryogenic distillation.

### Microchannel Distillation: Process Intensification of Isotopic Separations [22]
Ward Tegrotenhuis – Pacific Northwest National Laboratory

Process intensification (PI) is an emerging field focused on reducing the size of chemical process equipment, such as distillation, which is particularly relevant for isotopic enrichment of noble gases including $^{39}$Ar and $^{136}$Xe. Micro channel distillation (MCD) is a new PI technology that has reduced the length of column needed for a separation stage to less than 0.5 cm compared to 2 cm for the best available laboratory structured packing from Sulzer. This reduces the column length for 10,000 separations stages from 200 meters to less than 50 meters. Furthermore, MCD columns can be operated horizontally, thereby mitigating the need for excessively tall columns. Data are presented for isotopic enrichment of methane, as well as other chemical separations, which are used in two case studies. The first case study is to produce 5 tons of 80% enriched $^{136}$Xe in 5–6 years for the next generation of $^{136}$Xe neutrinoless double-beta decay experiments. The second is enrichment of germane to produce 1 tonne of 86% $^{76}$Ge in 4 years for a future LEGEND neutrinoless double-beta decay experiment. In both cases, calculations were performed using estimated relative volatilities of isotopes which have not been measured. So far, 60 separation stages have been demonstrated in a single 25-cm-long device, and further development is required to improve performance, scale-up to larger devices, and to integrate systems.

## 3.2 Production of Argon Radioisotopes in the Environment

The natural production of argon radioisotopes is an important aspect to low-radioactivity underground argon. Understanding the mechanisms for producing $^{37}$Ar, $^{39}$Ar, and $^{42}$Ar underground in the geology where the argon is located aids in the search for new sources of low-radioactivity underground argon and for characterizing current and future sources. Once the underground argon has reached the surface of the earth, it is critical to be able to calculate the production of the argon radionuclides to establish the requirements for transporting and storing the low-radioactivity underground argon.

### Subterranean production of argon-39 and implications for Doe Canyon well gas [24]
Ondřej Šrámek – Charles University

Low-radioactivity argon sources are desired by the WIMP dark matter experimental particle physics community. Accurate understanding of the subsurface production rate of the radionuclide $^{39}$Ar is also necessary for argon dating techniques and noble gas geochemistry of the shallow and the deep Earth.

Our new calculations of subsurface production of neutrons, $^{21}$Ne, and $^{39}$Ar [25] take advantage of the state-of-the-art reliable tools of nuclear physics to obtain reaction cross sections and spectra (TALYS) and to evaluate neutron propagation in rock (MCNP6). We discuss our method and results in relation to previous studies and show the relative importance of various neutron, $^{21}$Ne, and $^{39}$Ar nucleogenic production channels. Uncertainty in nuclear reaction cross sections, which is the major contributor to overall calculation uncertainty, is estimated from variability in existing experimental and library data. Depending on selected rock composition, on the order of $10^7$–$10^{10}$ alpha particles are produced in one kilogram of rock per year (order of 1–10 kg$^{-1}$ s$^{-1}$); the number of produced neutrons is 6 orders of magnitude lower, the $^{21}$Ne production rate drops by an additional factor of 15–20, and another one order of magnitude or more is dropped in production of $^{39}$Ar. Calculated $^{39}$Ar production rates span a great range from 29 ± 9 atoms per kg-rock per year in the K–Th–U-enriched Upper Continental Crust to (2.6 ± 0.8) × $10^{-4}$ atoms per kg-rock per year in the Depleted Upper Mantle. Nucleogenic $^{39}$Ar production exceeds the cosmogenic production below ~700 m depth in the Earth.

Recent report by the DarkSide-50 Collaboration [1](Agnes et al., 2016) puts the concentration of $^{39}$Ar in Doe Canyon, SW Colorado, deep $CO_2$ well gas at 1400±200 times lower compared to the atmospheric value. While it was argued that the Doe Canyon gas is derived from the Earth's upper mantle, it shows some counterintuitive isotopic characteristics, such as extremely low $^3$He/$^4$He suggesting crustal origin, while at the same time extremely low $^{39}$Ar/$^{40}$Ar indicating a source low in K, Th, U abundances. As a possible solution to this puzzle, we envisage a sizeable (i.e., low surface to volume ratio) gas reservoir in the shallow crust where mantle gas, contaminated by crust-derived gases ($^4$He, $^{21}$Ne, and $N_2$), accumulates for sufficient time (> 104 years) so that the $^{39}$Ar activity drops to the observed low value.

Overall, the noble gas observations (esp. of helium, neon, argon), both in terms of outgassing rates from the Earth and the gas' isotopic composition, present a major challenge to geoscientists who strive to formulate a coherent story of the Earth's formation, evolution, and current state. A link between underground noble gas production and decay of long-lived radionuclides ($^{40}$K, $^{232}$Th, $^{238}$U) ties noble gas geochemistry to dynamic models of thermal evolution of the Earth and to questions about deep Earth's composition and architecture.

This work was supported by the Czech Science Foundation under grant no. GAČR 17-01464S

### Background sources of argon-37 [26]
Christine Johnson – Pacific Northwest National Laboratory

Argon-37 is a radioactive isotope of argon that is naturally produced in both the atmosphere and the subsurface. It has a half-life of 35.04 days and decays via electron capture, emitting Auger electrons and x-rays as it decays. In both the atmosphere and the shallow subsurface, $^{37}$Ar is produced primarily by cosmic neutron reactions, with the highest production rate occurring at approximately two meters underground. The shallow subsurface production rate of $^{37}$Ar is dependent primarily on geology, but other factors such as location (latitude), altitude, and the solar cycle can also have an impact on the production rate. Many measurements have been made of the $^{37}$Ar concentration in both atmospheric and shallow subsurface air and a selection of these will be presented. Additionally, potential production mechanisms deep underground will be examined.

### Cosmogenic isotope production in argon [27]
Richard Saldanha – Pacific Northwest National Laboratory

Ar-39, a primarily cosmogenic isotope produced by neutron interactions with argon in the atmosphere, is the dominant background for argon-based dark matter detectors. It has been demonstrated that argon gas derived from underground sources has significantly reduced concentrations of $^{39}$Ar compared to atmospheric argon. To be used in dark matter detectors, gas extracted from underground must undergo substantial processing and purification in order to reach the chemical purity required. Cosmogenic activation of the underground argon during this processing is a potential concern for dark matter experiments.

In this talk we will review the current best estimates of the cosmogenic activation rates of $^{39}$Ar in argon, highlighting the lack of experimental measurements and the wide range of predicted cross sections. We will describe our experimental effort (currently underway) to measure the activation rate using a high energy neutron beam at the Los Alamos Neutron Science Center (LANSCE) and subsequent measurement of the $^{39}$Ar in ultra-low background proportional counters at PNNL. We will also briefly discuss other cosmogenic isotopes that are of interest to low-radioactivity argon-based detectors.

## 3.3 Uses of Argon Radioisotopes in the Environment

Due to the relatively long half-lives, argon radioisotopes are used as tools in environmental sciences. Ground water age-dating uses the fixed concentration of $^{39}$Ar in the atmosphere to discern groundwater movement in the earth to get a better understanding of our most precious resource. Argon-37 produced in the ground during an underground nuclear explosion can be used to monitor for such tests; helping to enforce international treaties. The challenges associated with these endeavors are closely related to the challenges with sourcing and characterizing low-radioactivity underground argon.

### Underground production of Ar-39: A field-data based assessment [28]
Roland Purtschert – University of Bern

With a half-life of 269 years Ar-39 is the best suited tracer for groundwater dating on age scales between 50-1000 years. It has been applied over the last 20 years in numerous studies and has proven to be a powerful tool for many groundwater studies. The interpretation of Ar-39 concentrations in terms of groundwater residence time is, in the ideal case, straight- forward. However, a potential limitation of Ar-39 dating of groundwater, is the possibility of underground production due to neutron activation of potassium. Over-modern Ar-39 concentrations have for example been observed in U- and Th- rich crystalline rocks. Also the concentrations of the short living Ar-37 (half-life 35 d) were very high in these cases, which indicates the usefulness of Ar-37 as a proxy for underground production. Another important factor is the escape path and probability of Ar-39 from the rock matrix into the water permeable pore space. In this review paper the importance of underground production of Ar-39 is assessed. Thereby Ar-39 and Ar-37 data collected over the last decade in numerous porous and fractured aquifers worldwide are discussed.

## Understanding $^{39}$Ar groundwater age dating: Importance in water resources protection and to identify potential alternative sources of low-radioactivity argon [29]

Signe White – Pacific Northwest National Laboratory

The ability to detect the level of depletion of $^{39}$Ar in groundwater relative to the modern atmospheric abundance provides valuable opportunities to determine groundwater age distributions and increase our understanding of groundwater systems. As an intermediate age tracer, $^{39}$Ar provides better constraints on groundwater age distributions than those determined from young and old age tracers alone ($^{3}$H, $^{14}$C, etc.). Typically, groundwater used for drinking and irrigation in many communities is from shallow, freshwater aquifer systems. Mixing of young and old water occurs in these systems and can introduce contaminants from the surface to precious groundwater resources. Knowing the age of the groundwater can also indicate areas where water resources are being depleted at a much faster rate than it is being replenished. Therefore, being able to determine groundwater age distributions is critical to protect our most valuable resource.

Groundwater increases in salinity with depth and becomes unsuitable for drinking without costly treatment. However, deep groundwater is typically older in age and very old water will be depleted in $^{39}$Ar, as long as there is no in situ production. In situ production of $^{39}$Ar typically takes place in groundwater areas where uranium, potassium, and thorium are abundant. Therefore, to find subsurface sources of argon that are depleted in $^{39}$Ar, we must look to identify deep hydrologic environments with very old water and with rocks that have a relatively low abundance of uranium, potassium and thorium. Knowledge gained in better understanding groundwater age distributions will in turn support identification of these areas.

This talk will discuss the importance of understanding groundwater age to support both environmental applications and the search for low-radioactivity argon. We will provide an overview of PNNL's application of $^{39}$Ar age dating of groundwater to support water resources studies, and briefly describe the types of groundwater systems that may be suitable sources of argon to support fundamental physics experiments.

## An Overview of Nuclear Explosion Monitoring at PNNL [30]

Michael Foxe – Pacific Northwest National Laboratory

Pacific Northwest National Laboratory has a long history of environmental monitoring. In support of this effort, PNNL performs a suite of Nuclear Explosion Monitoring research and development. This R&D focuses on gas processing systems and nuclear detectors with an emphasis on radioxenon and radioargon measurements. In this presentation, an overview of the NEMP program will be provided along with a description of recently developed radioxenon and radioargon detection systems and the respective capabilities.

## 3.4 Measuring Low-Levels of Argon Radioisotopes

Measuring low-levels of $^{39}$Ar and $^{37}$Ar have largely been driven by the environmental sciences; some of which were discussed in the previous section (3.3). However, due to the stringent requirements on the $^{39}$Ar level, the dark matter physics community is also developing new technologies to assay low-radioactivity underground argon for $^{39}$Ar. While $^{37}$Ar is measured strictly through radioactive decay, $^{39}$Ar is measured through accelerator mass spectrometry, atom trap trace analysis (ATTA), and also through radioactive decay. Argon-42 is also an important isotope for some of the physics endeavors, and although it has been measured in the atmosphere, there is no method for routine assay of argon for $^{42}$Ar.

## Detection of $^{39}$Ar using positive ion Accelerator Mass Spectrometry - an overview [31]
Philippe Collon – University of Notre Dame

The first application of $^{39}$Ar Accelerator Mass Spectrometry (AMS) at Argonne National Laboratory was to date ocean water samples relevant to oceanographic studies using the gas-filled magnet technique to separate the $^{39}$K-$^{39}$Ar isobars. In particular the use of a quartz liner in the plasma chamber of the electron cyclotron resonance ion source enabled a $^{39}$K reduction of a factor ~130 compared to previous runs without liners, and allowed us to reach a detection sensitivity of $^{39}$Ar/Ar = $4.2 \times 10^{-17}$. In order to improve this sensitivity and allow the measurement of lower ratios, higher ion source currents and a lower overall $^{39}$K background environments needed to be developed.

The talk will give an overview of our efforts to detect low concentrations of $^{39}$Ar as well as efforts to investigate new methods combining low-level potassium cleaning techniques with the use of ultra-pure aluminum liners in the plasma chamber of the ion source, with the aim of gaining the 1 to 2 orders of magnitude in $^{39}$Ar detection sensitivity required in the selection of ultra-pure materials for detectors used in weakly interacting massive particle dark matter searches.

## Atom Trap Trace Analysis of Rare Noble Gas Isotopes [32]
Peter Mueller – Argonne National Laboratory

Atom Trap Trace Analysis (ATTA) is an efficient and highly selective laser-based atom counting method where neutral atoms of the desired isotope are captured by laser light in a magneto-optical trap and detected one-by-one via laser induced fluorescence. ATTA is unique among trace analysis techniques in that it is free of interferences from any other isotopes, isobars, atomic or molecular species. At Argonne, we have brought this novel technique from a proof-of-principle concept all the way to a routine analytical tool that is now available at large to the Earth science community and beyond. Currently, we concentrate on two long-live krypton isotopes: Kr-81 (half-life = 230,000 yrs) as an ideal tracer for old water and ice with mean residence times in the range of $10^5$–$10^6$ years, and the anthropogenic isotope Kr-85 (half-life = 10.8 yrs) for dating respectively young ground waters. Both isotopes can now be routinely analyzed via ATTA in samples as small as 5 microliters of krypton gas (STP). More generally, other rare noble gas isotopes such as Ar-39 and He-3 are amenable to ATTA analysis, and in particular, ATTA Ar-39 instruments are currently under development at the University of Heidelberg and the University of Science and Technology of China. I will present the current state-of-the-art of the ATTA technique, examples of recent applications, and how this technology can relate to low-radioactivity noble gas materials. This work is supported by Department of Energy, Office of Nuclear Physics, under Contract No. DEAC02-06CH11357.

## Measuring the $^{39}$Ar depletion factor with DART [33]
Pablo García - CIEMAT

The DarkSide-20k (DS-20k) Dark Matter search experiment will operate with 37-ton radio-pure underground argon (UAr), extracted from the Urania plant in Cortez (USA) and purified in the Aria distillation plant (Sardinia, Italy). The $^{39}$Ar depletion factor in UAr with respect to atmospheric argon is expected to be around 1400. Assessing the purity of UAr in terms of $^{39}$Ar is key for the physics programme of DS-20k.

DART is a small (~1 L) chamber that will measure the depletion factor of $^{39}$Ar in UAr. The detector will be immersed in the LAr active volume of ArDM (LSC, Spain), which will act as a veto for gammas stemming from the detector materials and from the surrounding rock. DART will use the SiPMs constructed for DS-20k and the cryogenic and DAQ systems of ArDM. Data taking is planned for 2018. In this talk, I will review the status of the DART project, which comprises the chamber design and construction, as well as the background studies necessary to assess the expected performance of DART.

## Sensitivity and detection limits for measuring low levels of argon radioisotopes ($^{37}$Ar and $^{39}$Ar) using ultra-low-background proportional counters [34]
*Emily Mace – Pacific Northwest National Laboratory*

Pacific Northwest National Laboratory (PNNL) has developed ultra-low-background detectors that enable the measurement of low levels of radioactivity for various applications such as $^{39}$Ar for age dating groundwater aquifers and $^{37}$Ar for detection of underground nuclear testing. These low-background detectors can achieve even greater levels of sensitivity when paired with the low-background counting system located in PNNL's Shallow Underground Laboratory (at ~35 meters water equivalent).

Argon-37 is a challenging isotope to measure due to the radioactive decay via Auger electrons and x-rays at very low energies (mean peak energy of 2.82 keV) and benefits from the use of an internal-source proportional counter for detection and quantification. Argon-39 is less of a challenge to measure because $^{39}$Ar is created in the atmosphere in abundance; it undergoes beta decay with an end-point energy at 565 keV. The greater challenge is to quantify a depletion of $^{39}$Ar relative to the modern atmospheric abundance.

This talk will review the current PNNL low-background capabilities that enable the detection of low levels of radioactivity from argon samples ($^{37}$Ar and $^{39}$Ar). We also present a methodology for using $^{39}$Ar to age-date groundwater aquifers and discuss the current detection limits and sensitivity to measure $^{37}$Ar and $^{39}$Ar using low-background proportional counters at PNNL.

## The Heidelberg ArTTA: Application of quantum technology for single atom $^{39}$Ar detection [35]
*Zhongyi Feng - Kirchhoff-Institut für Physik, Universität Heidelberg*

Argon Trap Trace Analysis (ArTTA) utilises established techniques from the field of atomic and optical physics for single $^{39}$Ar detection. It exploits the shift of the optical resonance frequency due to differences in mass and nuclear spin to detect this rare isotope down to the $10^{-16}$ level. Although a single resonant excitation is not sufficient to distinguish between isotopes, many cycles of photon absorption and emission guarantee perfect selectivity. Compared to low-level counting (LLC) and accelerator mass spectrometry (AMS), our method is neither dependent on radioactivity nor affected by isobars. In this talk, we present results of an $^{39}$Ar intercomparison study between the LLC laboratories from PNNL and Bern with our Heidelberg ArTTA collaboration. We further present the current status of our apparatus and demonstrate the prospects for measurements on a routine basis with an exemplary oceanographic study. The measurements have been performed with a typical sample size of 2 mL STP argon corresponding to 5 L of ocean water. These applications show the versatile potential of ArTTA.

## 3.5 High-Level $^{39}$Ar Applications

Although the workshop was focused on argon with very low levels of argon radionuclides, high levels of $^{39}$Ar have direct application in the sciences that low-radioactivity underground argon enables. A supply of argon with high-levels of $^{39}$Ar allows for the production of low-level $^{39}$Ar standards for precise detector calibration. Precise measurement of the $^{39}$Ar beta-decay spectrum and an understanding of detector limitations (e.g., pulse-shape discrimination efficacy) are enabled by large datasets of $^{39}$Ar decays.

## Development of a Low-Level $^{39}$Ar Calibration Standard - Quantified by Absolute Gas Counting Measurements [36]
Richard Williams – Pacific Northwest National Laboratory

Argon-39 is an attractive environmental radiotracer for monitoring phenomena related to ground-water transport. With a half-life of 269 years, $^{39}$Ar provides continuity and overlap with other more commonly used radiotracers such as $^3$H and $^{14}$C. The age of a water sample is determined by comparing the $^{39}$Ar specific activity of the gas separated from ground-water with equilibrium $^{39}$Ar atmospheric levels (1.01 Bq/kg-Ar$_{NAT}$ or 1.80×10$^{-6}$ Bq/cc-Ar$_{STP}$ ). Such measurements require the use of low-level gas counting techniques which have been calibrated using well characterized $^{39}$Ar standards. This presentation describes the generation of $^{39}$Ar, via reactor irradiation of potassium carbonate, followed by quantitative analysis, based on length-compensated proportional counting, to yield standards that are approximately 60 and 3 times atmospheric background levels of $^{39}$Ar. Multiple measurements of the 60× standard, at various pressures, were performed in Pacific Northwest National Laboratory's shallow underground counting laboratory in order to study the effect of gas density on beta-transport within the counters. In order to estimate the specific activity of the standard from measurements based on length-compensated proportional counting it is necessary to account for disintegrations that do not deposit sufficient energy in the counter to register above threshold. These losses are commonly referred to as the so-called Wall-Effect (betas reaching the wall before depositing sufficient energy for detection) and Threshold-Effect (total deposited energy below detection threshold). For this study both the Wall- and Threshold-Effect have been estimated using Monte-Carlo simulations and applied to the experimental measurements. An uncertainty model of the measurements and data analysis has been developed in accordance to the Guide to the Expression of Uncertainty in Measurements (GUM). The most challenging source of uncertainty to quantify is that from the Monte-Carlo simulations. The total expanded uncertainty (K=2) result for the 60×-background $^{39}$Ar standard, less any uncertainty contribution from the Monte-Carlo simulation, is 1.3% (approximately 95% confidence). Efforts to estimate the magnitude of the uncertainty from the simulation are discussed along with future directions for improved simulations.

## Measurement of the $^{39}$Ar β-Spectrum in Natural Argon with the DEAP-3600 Experiment [37]
Pietro Giampa – TRIUMF

The DEAP-3600 experiment is a single-phase liquid argon detector, located 2 km underground at the SNOLAB facility in Sudbury, ON, Canada. The detector, capable of holding up to 3.6 tonnes of natural liquid argon, is designed and optimized to perform a WIMP Dark Matter search with a projected sensitivity to the spin-independent WIMP-nucleon cross-section of 10$^{-46}$ cm$^2$ (for a 100 GeV WIMP). Complementary to its main purpose, the DEAP-3600 experiment can be also used to perform precision measurements on the cosmic-ray activated $^{39}$Ar isotope. In this talk, we will report on the β-spectrum extraction, and how this was utilized for the energy calibration of the experiment. Moreover, we will present the estimated $^{39}$Ar activity in natural argon, from the DEAP-3600 dataset. Finally, we will also cover how precision measurements of the $^{39}$Ar β-spectrum can be used as the building block for beyond-WIMP Dark Matter searches.

## High activity argon-39 source measurements [38]
Christopher Jackson – Pacific Northwest National Laboratory

Argon-39 is a significant background in current and future dark matter detectors (depending on the eventual reduction level in underground sources). Measurements to understand the spectral shape, half-life and decay properties, as well as the limits of pulse-shape discrimination, will be important to understand future sensitivity to interesting physics. In this talk the creation and initial measurement of a high activity argon-39 source and possible future uses are described.

# 4 Acknowledgements

Special thanks to the external advisory committee: Walter Bonivento, Mark Boulay, Steven Elliott, Cristian Galbiati, Art McDonald, and Andrew Renshaw. Thank you to the efforts of the local workshop organizing committee and administrative help: Craig Aalseth, Raymond Bunker, Ben Loer, Yolanda Olivera, Melissa Bang, and Blake Wright.

The Low-Radioactivity Underground Argon workshop was made possible by the Nuclear-physics, Particle-physics, Astrophysics, and Cosmology (NPAC) Initiative, a Laboratory Directed Research and Development (LDRD) Program at Pacific Northwest National Laboratory (PNNL). PNNL is a multi-program national laboratory operated by Battelle Memorial Institute for the U.S. Department of Energy under Contract No. DE-AC05-76RL01830.